\documentclass[onecolumn]{aastex62}

\pdfoutput=1 
\usepackage{amsmath,amstext}
\usepackage[T1]{fontenc}
\usepackage{apjfonts} 
\usepackage[figure,figure*]{hypcap}
\usepackage{subfigure}
\usepackage{epsfig}
\usepackage[utf8]{inputenc}
\usepackage{longtable}
\usepackage{isotope}

\newcommand{\cutt}[1]{\textcolor{blue}{}}

\newcommand{\Ms}{{\ensuremath{\text{M}_{\odot}}}}
\newcommand{\GIZMO}{\texttt{GIZMO}}
\newcommand{\TNG}{\texttt{IllustrisTNG}}
\newcommand{\Mach}{$\mathcal{M}$}
\newcommand{\NumDen}{cm$^{-3}$}

\shorttitle{Supersonic Turbulence in Minihalos}

\shortauthors{Ho, Chen, and Tung}

\begin{document}

\title{Turbulence in Primordial Dark Matter Halos and Its Impact on the First Star Formation}

\author[0000-0002-2316-1371]
{Meng-Yuan Ho}

\affiliation{Institute of Astronomy and Astrophysics, Academia Sinica, Taipei 10617, Taiwan} 
\affiliation{Department of Physics, National Taiwan University, Taipei 10617, Taiwan} 
\affiliation{Department of Physics and Astronomy, University of California, Riverside 92521, California, USA} 

\author[0000-0002-4848-5508]{Ke-Jung Chen}
\affiliation{Institute of Astronomy and Astrophysics, Academia Sinica, Taipei 10617, Taiwan} 
\affiliation{Heidelberg Institute for Theoretical Studies, Schloss-Wolfsbrunnenweg 35, 
Heidelberg 69118,
Germany}

\author[0009-0000-9401-5470]
{Pei-Cheng Tung}
\affiliation{Institute of Astronomy and Astrophysics, Academia Sinica, Taipei 10617, Taiwan} 
\affiliation{Department of Physics, National Taiwan University, Taipei 10617, Taiwan}

\begin{abstract}

We present high-resolution simulations of the first star-forming clouds in 15 minihalos with masses ranging from $\sim 10^5$ to $10^7\ \Ms$ at redshifts $z \sim 17$--$20$, using the \texttt{GIZMO} code. Our simulations incorporate detailed primordial gas physics and adopt initial conditions from the state-of-the-art \TNG\ cosmological simulations. To achieve the required resolution, we apply a particle-splitting technique that increases the resolution of the original \TNG\ data by a factor of $\sim 10^5$, reaching gas and dark matter particle masses of $0.2\ \Ms$ and $80\ \Ms$, respectively. This enables us to resolve gas accretion during the early assembly of minihalos and to capture the emergence of strong turbulent flows. We find that turbulence, driven by gas infall into the dark matter potential wells, is predominantly supersonic, with characteristic Mach numbers ranging from $1.8$ to $4.2$, increasing with halo mass. The supersonic turbulence effectively fragments the central gas cloud into multiple clumps. Some of dense clump masses range from $2.6~\Ms$ to $66.5~\Ms$, exceeding their corresponding Jeans masses and soon collapsing to form the first stars. Our results suggest that supersonic turbulence is a common feature in minihalos and plays a key role in generating clumpy star-forming clouds, with important implications for the initial mass function of the first stars.

\end{abstract}


\section{Introduction}

The first generation of stars, known as Population III (Pop~III) stars, formed from pristine primordial gas composed primarily of hydrogen and helium. These stars are believed to have formed in dark matter halos with masses of $10^5 - 10^6$\Ms\ \citep{Tegmark_1997, Bromm_2002, Norman_2008}. In the absence of metals, the only efficient coolant in these early clouds at temperatures below $10^4$ K is molecular hydrogen, which enables gas to cool down to $\sim200$ K and initiates gravitational collapse \citep{Yoshida_2003_CoolingH2, Stiavelli_2009}. Unlike present-day star formation, the lack of metal-line cooling in primordial environments results in significantly higher Jeans masses, leading to the formation of more massive stars \citep{Abel_2002, Yoshida_2003_CoolingH2}. Early theoretical studies suggest that Pop~III stars formed with masses ranging from 40 to 500 \Ms, far exceeding the typical stellar masses observed in the local universe \citep{Tegmark_1997, abel1998formationfragmentationprimordialmolecular_PopIII_0, 2001ApJ...546..635O_PopIII_1, Omukai_2003_PopIII_1.5, 2007ApJ...654...66O_PopIII_2, Hirano_2014_PopIII_3, Hosokawa_2016_PopIII_4, klessen2023starsformationpropertiesimpact_PopIII_5}.  

Recent simulations predict a broad range of Pop III stellar masses. \cite{Lake_2025} find IMFs spanning $\sim 0.1 - 100$ \Ms, both with and without jet feedback. \cite{wollenberg20}  report a protostellar mass function extending from $\sim 10^{-3}$ to 10 \Ms\ in the presence of turbulence. Using sink particles, \cite{jaura22} obtain mass distributions from $\sim 10^{-2}$ to 100 \Ms, with turbulent models producing a larger number of low-mass objects. \cite{Prole_2022} find IMFs ranging from $\sim 10^{-2}$ to 20 \Ms, depending on the sink-particle and sub-grid turbulence receipts. Overall, these studies indicate that the mass distribution of Pop III protostars in the early Universe spans approximately $10^{-3} - 10^2$ \Ms, with the exact range depending on the adopted physical assumptions, numerical methods, and the inclusion receipts of turbulence and feedback processes. While the Pop III star-forming cloud holds the key to determine the IMF of Pop III, the aim of this study is to better understand the physical properties of the Pop III star-forming cloud in the mini halo which can be better modeled and less affected by the SF receipts adopted in various simulations.

Furthermore, observations of extremely metal-poor (EMP) stars—believed to be second-generation stars enriched by Pop~III supernovae—indicate that their progenitors may have had lower masses, in the range of $25$–$50\ \Ms$ \citep{Umeda_2005_EMP_1, Ishigaki_2018_EMP_2, 2024A&A...681A..44S_EMP_3, 2024ApJ...961L..41J_EMP_4}. The formation of Pop~III stars is influenced by several physical processes, including turbulence, radiative feedback, and magnetic fields \citep{Menon_2020_2, Menon_2020, Mathew_2021, Grudi_2022, He_2023, Sharda_2024}. We propose that the apparent discrepancy between theoretical predictions and observational constraints may stem from the limited resolution in previous simulations, which often failed to capture the emergence and effects of turbulence during minihalo assembly \citep{turk09,stacy10,hir13,Greif_2015,chen15}. These previous studies used hierarchical zoom-in techniques to resolve the central star-forming regions with high-resolution of sub-pc scale, but the larger-scale environment of the halo of $100-1000$ pc remained poorly resolved, making it challenging to model the emergence of turbulence from gas accretion during halo assembly.

The recent study by \citet{2025_chen} was the first to demonstrate the presence of supersonic turbulence emerging during the formation of minihalos. Their simulations revealed that this turbulence can drive fragmentation at the cloud scale, potentially influencing both the subsequent star formation and the characteristic mass scale of Pop~III stars. However, \citet{2025_chen} focused on a single, massive minihalo of $\sim 10^7\,\Ms$, leaving the general nature and prevalence of turbulence in minihalos largely unexplored. 
In this study, we extend \cite{2025_chen} to a broader sample of halos to investigate the diversity of turbulence in the early universe. We select a suite of minihalos from the large-scale cosmological simulation \TNG, and enhance the resolution by a factor of $10^5$ through a particle-splitting technique. This enables us to resolve the gas accretion onto halos and capture the development of turbulence within primordial gas clouds. This paper aims to characterize the physical properties of turbulent gas, examine its dependence on host halo properties, and assess its role in shaping the initial conditions for Pop~III star formation.

The structure of this paper is as follows. In Section 2, we describe the \GIZMO\ code, including the method used to initialize conditions from the \TNG\ simulations and the implementation of the particle-splitting algorithm. In Section 3, we present the formation of supersonic turbulence in minihalos and examine the physical properties of clumpy star-forming regions. Section 4 discusses the astrophysical implications of our findings, and we conclude in Section 5.

\section{Methods}

In this section, we describe the details of our simulation setup. The initial conditions are derived from the \TNG\ project, a state-of-the-art large-scale cosmological simulation that captures the hierarchical growth of structure and the evolution of baryons and dark matter across cosmic time. From the \TNG\ snapshot at $z \approx 20$, we identify and select 15 minihalos with masses in the range of $10^5$--$10^7 \Ms$, representing the typical environments for the Pop~III star formation. These halos are then extracted from \TNG\ and re-simulated using the \GIZMO\ code in meshless finite mass (MFM) mode, which allows for accurate modeling of gas dynamics in complex astrophysical environments.

Since the simulation resolution of the \TNG\ is insufficient to resolve the small-scale gas dynamics within these minihalos, we employ a particle-splitting algorithm to increase the resolution by a factor of $10^5$. This procedure significantly refines the mass resolution of gas particles, enabling us to resolve the turbulent gas accretion and fragmentation processes that occur during the assembly of minihalo.

\subsection{GIZMO}

\GIZMO, developed by \citet{Hopkins_2015}, is a robust hydrodynamics code based on the Meshless Finite Mass (MFM) and Meshless Finite Volume (MFV) methods. It evolves from the widely used \texttt{GADGET-2} code \citep{Gadget-2-2005MNRAS.364.1105S}, originally designed for N-body/SPH cosmological simulations. Like \texttt{GADGET-2}, \GIZMO\ employs a Tree Particle-Mesh (TreePM) algorithm \citep{Xu1995, Bagla2002, Bode2003, Hopkins_2015} for gravitational calculations, enabling accurate and efficient force evaluations by hierarchically decomposing the simulation volume.

The MFM/MFV approach in \GIZMO\ replaces the traditional smoothed-particle hydrodynamics (SPH) kernel summation with a volume-discretized formulation that combines the strengths of both SPH and grid-based Eulerian methods. This hybrid formulation allows for superior accuracy in capturing shocks, contact discontinuities, and turbulent flows while maintaining excellent conservation properties and numerical stability. \GIZMO\ has been rigorously validated in a wide range of astrophysical and cosmological test problems, establishing its reliability for high-resolution simulations.

\subsubsection{Physics of Primordial Gas}

We model the cooling and chemistry of primordial gas in our \GIZMO\ simulations using \texttt{GRACKLE} \citep{2017MNRAS.466.2217S_Grackle}, a publicly available chemistry and cooling library. \texttt{GRACKLE} tracks a non-equilibrium network of twelve primordial species: $\mathrm{H,\, H^+,\, H^-,\, D,\, D^+,\, HD,\, H_2,\, H_2^+,\, He,\, He^+,\, He^{++},\, e^-}$. The associated gas cooling processes include collisional excitation and ionization, radiative recombination, free-free emission, and other atomic cooling mechanisms for hydrogen and helium. Additionally, \texttt{GRACKLE} accounts for molecular cooling from $\mathrm{H_2}$ and HD through rotational and vibrational line emission, formation heating/cooling, and collision-induced emission. The chemistry and cooling network is self-consistently coupled with the hydrodynamic evolution, allowing for accurate modeling of the thermal and chemical state of primordial gas.

\subsection{Mini-Halos in the \TNG}

The \TNG\ project is a state-of-the-art suite of large-scale cosmological magnetohydrodynamical simulations designed to study galaxy formation and evolution from redshift $z = 127$ to $z = 0$ \citep{2018Nelson, 2018Naiman, 2018Springel, 2018Marinacci}. These simulations are performed using the moving-mesh code \texttt{AREPO} \citep{2010Springel, 2016Weinberger, 2018Pillepich} and cover a wide range of cosmic volumes, mass resolutions, and baryon physics. For this study, we utilize the highest-resolution dataset from the TNG50-1 run, which is particularly suitable for modeling minihalos. The baryonic mass resolution in TNG50-1 is approximately $8.5 \times 10^4\,\Ms$, and the dark matter mass resolution is about $4.5 \times 10^5\,\Ms$.

We identify minihalo candidates using the \texttt{SUBFIND} algorithm \citep{2001Springel} through the publicly available \TNG\ data release\footnote{\url{https://www.tng-project.org/data/}}. Our selection criteria are as follows: we search for halos with total masses of $10^5-10^7\,\Ms$ in the earliest snapshot at $z = 20.05$. We further restrict our selection to halos that are non-star-forming and relatively isolated, excluding those that undergo major mergers between $z \approx 20$ and $z \approx 15$. 

In IllustrisTNG, the selected halos correspond to overdense regions relative to the cosmic mean and are embedded in a gaseous environment. Owing to the limited resolution of the original IllustrisTNG initial conditions, small-scale structures such as filaments cannot be clearly identified. Moreover, at $z\sim20$, large-scale cosmic structures are still only weakly developed. As a result, the selected halo progenitors are relatively isolated, with no star-forming systems within a radius of 10 comoving kpc.
The halo progenitors are identified as subhalos, which may themselves reside within more massive host halos. In cases where the free-fall time of the host halo is much longer than the duration of our simulations, we treat the progenitor region as effectively isolated and unaffected by the evolution of the larger-scale environment. In our halo-selection procedure, we require that neither the subhalo nor its host halo contains any star particles. We further ensure that the selected volume is completely free of star particles, which provides a sufficient and robust criterion for isolating pristine halo progenitors in our simulations.

Based on these criteria, we select 15 minihalo candidates of total masses in the range $4.9 \times 10^5$ to $7.7 \times 10^6\,\Ms$. For each halo, we extract a comoving spherical volume of radius $\sim 10$~ckpc from \TNG\ that encompasses the halo and its surrounding medium. This volume is then mapped directly onto our \texttt{GIZMO} simulations as the zoom-in initial conditions. 

\subsection{Cosmological Zoom-in Simulations via particle-splitting}

These TNG halos originally contained only tens to hundreds of gas and dark matter particles. Such a mass resolution is too low to resolve the structure of the gas flow. To further increase the mass resolution of the initial conditions of TNG50-1, we apply the super-Lagrangian refinement introduced in \cite{Hopkins_2015}, an algorithm for splitting particles to increase the mass resolution of the original TNG50-1 halos. This method is widely used in zoom-in simulations, e.g. \cite{2024Tung,2024Ramesh_1,2024Ramesh_2}. 

To resolve the turbulence produced in the simulation, we need to increase the mass resolution of approximately $8.38 \times 10^4\,\Ms \rightarrow 0.19\,\Ms$ for gas particles and $4.56 \times 10^5\,\Ms \rightarrow 80.88\,\Ms$ for dark-matter particles. The particle splitting algorithm operates as the following steps:

\begin{enumerate}
    \item \textbf{Criterion Check for Splitting} Determine if a particle meets the splitting criteria. Specifically, a particle is eligible for splitting if its mass exceeds two times the targeted masses, which are 0.19 \Ms~for gas and 80.88 \Ms~for dark matter.
    
    \item \textbf{Determining Spacing of child particles} Before splitting a particle, determine the spacing of the new particles. The code searches for the nearest particle to the target particle and then calculates the effective radius on the basis of the distance between them or the softening length. The spacing distance is set to the minimum of the two values: 0.70 of the effective radius, which is determined by the radius to enclose 32 neighbors, or 0.25 of the softening length with a cubic spline kernel.
    
    \item \textbf{Splitting} If the particle satisfies the splitting criteria, it will be divided into two new particles, each with half the mass of the original particle. The positions of the two particles are randomly adjusted within a specified direction and separated by the spacing distance while maintaining the conservation of mass, momentum, and inertial momentum.

\end{enumerate}

Using this particle-splitting algorithm, we can achieve the desired mass resolution to resolve the turbulent gas during the formation of minihalos.
Particle splitting is applied during the first five million years of the simulation. In this phase, only the self-gravity of gas and dark matter, along with gas hydrodynamics, is active. The splitting proceeds gradually, progressively refining particle masses and enabling the resolution of evolving structures driven by gravitational collapse. By $5$~Myr, the particle masses reach the target resolution for both gas and dark matter, and dense gas structures begin to emerge. Unlike adaptive mesh refinement (AMR) methods, which refine only preselected regions of interest, our particle-splitting approach is applied uniformly to all particles within the simulation domain. This strategy ensures consistent mass resolution throughout the entire volume, providing a more complete picture of the developing turbulence.

In the particle splitting process, the mass ratio between the dark matter and gas particle is 10 which hold on until the mass of gas particles refines under 8 \Ms, corresponding to 80 \Ms\ for dark matter particles. The limitation of our computing capacity prevents us from going to a higher resolution of dark matter when the N-body calculation of dark matter becomes increasingly difficult to further evolve the simulations. Furthermore, the nature of dark matter is poorly understood, and whether it is reasonable to assume dark matter behaves like particles with such a high resolution. As a result, we stop splitting dark matter particles while the particle mass of DM falls below 80 \Ms.

After reaching the desired resolution, we activate the primordial gas cooling and non-equilibrium chemistry modules using the \texttt{GRACKLE} library, and continue to evolve the mini-halo. For the initial gas composition, we assume a primordial chemical composition for all gas particles, consisting of atomic hydrogen ($76\%$), helium ($24\%$), free electrons, and trace amounts of deuterium (D) and lithium (Li). The initial metallicity is set to a very low value of ($Z = 10^{-6}$\ $Z_\odot$), and the initial ($\mathrm{H}_2$) fraction is set to zero. During the particle-splitting phase, gas densities remain low, and ($\mathrm{H}_2$) formation is therefore inefficient. After particle splitting, as dense structures begin to form, we enable the \texttt{GRACKLE} chemistry network to follow the subsequent non-equilibrium chemical and thermal evolution of the gas.

Since our objective is to investigate the formation of turbulence during the gravitational assembly of minihalos without stellar feedback, we intentionally exclude any prescriptions for star formation. Toward the end of the simulations, the dense gas undergoes rapid collapse due to Jeans instability, which drastically reduces the simulation timestep due to the Courant–Friedrichs–Lewy (CFL) condition \citep{nick18_cfl}. This computational slowdown ultimately leads to the termination of the simulation. As a result, each model evolves over a slightly different duration, typically from $z = 20$ to $z = 18.96$–$18.26$, corresponding to 14.17–24.80 million years of physical time. For consistency, we used the final snapshot available from each model for our analysis. 

In the original IllustrisTNG project, subgrid models are used to treat physical processes below the resolution scale \citep{2018Nelson}. In this work, we instead focus on turbulence that arises purely from the gravitational potential of halos. We use TNG-50 to generate cosmologically realistic initial conditions for our zoom-in simulations, but do not include cosmic density fluctuations below the resolution of the parent Illustris simulation. This raises the question of whether the resulting gas structure is robust to increases in mass and spatial resolution based on the particle-splitting method. We therefore examine the effects of particle resolution in the following section.

\subsection{Resolution Tests}

The initial conditions of TNG do not resolve minihalo formation and do not explicitly include small-scale CDM perturbations. Nevertheless, TNG-50 is among the highest-resolution publicly available cosmological simulations, making it a suitable framework given its cosmological consistency and accessibility. We therefore adopt its initial conditions as a baseline and subsequently apply an iterative particle-splitting technique to increase the effective resolution, enabling us to follow the detailed physical processes during minihalo assembly. While small-scale CDM perturbations are not explicitly included, small perturbations are introduced to the split particles to mimic sub-particle–scale fluctuations.

In the IllustrisTNG initial conditions, the particle masses are $4.5 \times 10^5\,\Ms$ for dark matter and $8.5 \times 10^4\,\Ms$ for gas, corresponding to a mass ratio of $\sim$5. We note that mass ratios significantly larger than this can introduce numerical artifacts, including two-body heating and artificial fragmentation or clumping. During the particle-splitting procedure, we initially adopt a gas-to-dark-matter mass ratio of 1:10, setting the highest dark-matter resolution to $100\ \Ms$, and preserve this ratio until the gas mass resolution reaches $10\ \Ms$.

In the context of minihalo formation, dark matter primarily establishes the gravitational potential that enables gas accretion and subsequent star formation. We therefore aim to achieve the highest feasible mass resolution for both components, subject to numerical stability and computational constraints. Further reduction of the dark-matter particle mass leads to prohibitively small timesteps and a substantial slowdown of the simulations. Moreover, given the uncertain microphysical nature of dark matter, resolving it with extremely small particle masses is not necessarily physically motivated.

In our final configuration, we adopt a dark-matter particle mass of $\sim 80\ \Ms$ and a gas particle mass of $\sim 0.2\ \Ms$, corresponding to highest spatial resolutions of 10~pc for dark matter and 0.05~pc for gas. The enhanced gas resolution is required to capture accretion-driven turbulence and its cascade. In addition, within the central regions of the halo, gas self-gravity exceeds that of dark matter, providing further justification for prioritizing gas resolution over the dark-matter component.

\begin{figure}[tbh]
\centering
\includegraphics[width=0.6\columnwidth]{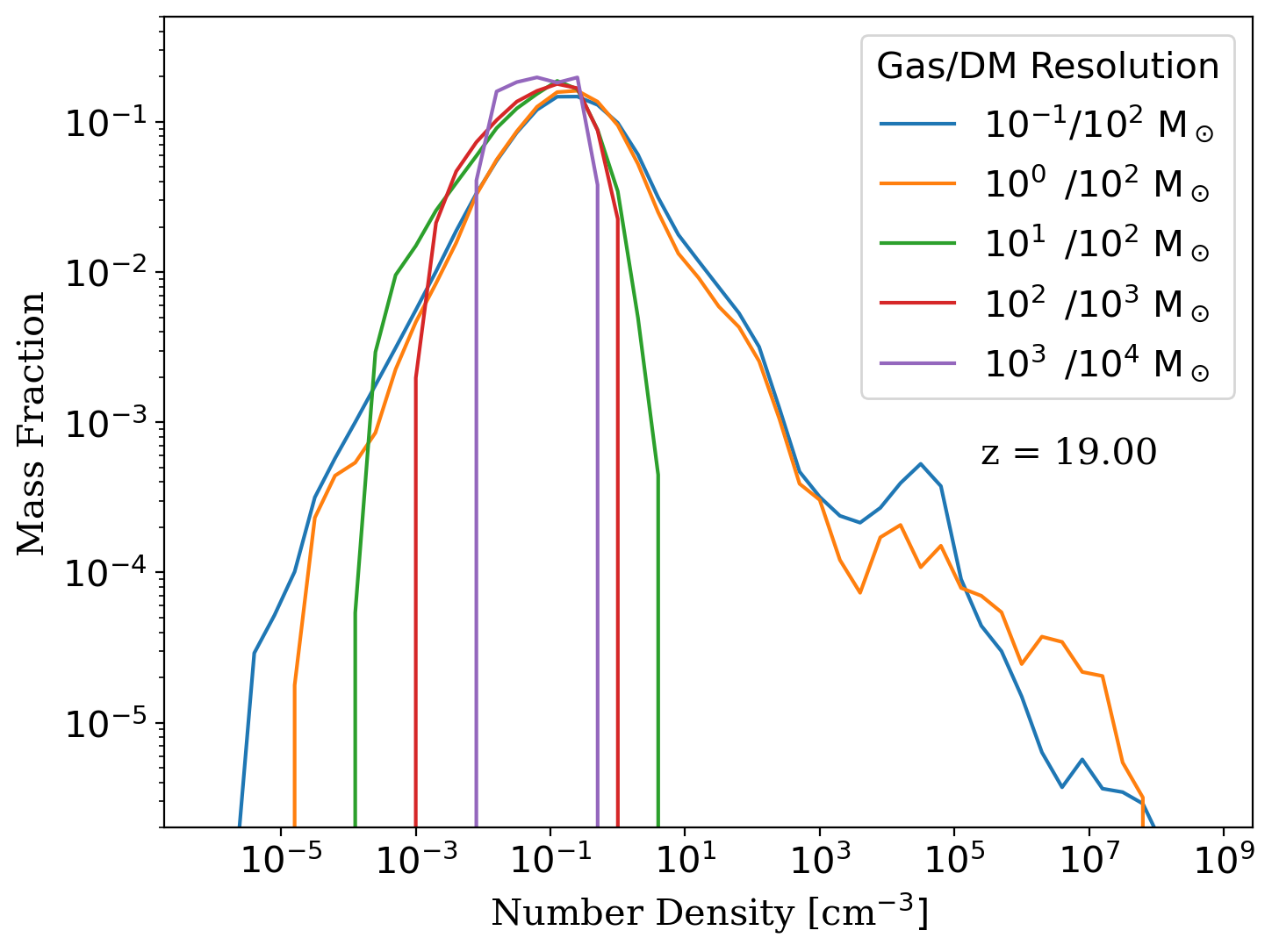}
\caption{Gas density distributions in a primordial halo. Profiles represent the results from the simulation with different mass resolutions. The peak of the distribution is consistent among different mass resolutions. Meanwhile, higher resolution extends to distribution in a broader density range. For the gas resolution of 0.1 and 1 \Ms, the profiles start to converge and their maximum density achieves up to $\sim 10^{8}$ \NumDen.}
\label{fig:res}
\end{figure}

To validate the effective resolution of our simulations, we modeled the formation of a $\sim 10^7\ \Ms$ minihalo using our methodology with gas mass resolutions of 100, 10, 1, and $0.1\ \Ms$. The resulting gas density distributions are shown in Figure~\ref{fig:res}. As the mass resolution increases, the maximum density increases and the minimum density decreases accordingly, while the peak of distribution remains located at the density of $\sim 10^{-1}$ \NumDen. When the mass resolution of gas jumps from 10 to 1 \Ms, the profiles become more extended because the upper and lower density shifts more. For a mass resolution of 1 \Ms, a tiny fraction of the gas can even reach a high density $> 10^{5}$ \NumDen, which normally refers to the gas-forming stars. When the mass resolution of the gas particles is reached at $1$ and $0.1$ \Ms, the distribution of density profiles starts to converge, reaching an effective resolution. Although the simulations cannot maintain a constant mass-resolution ratio between gas and dark matter at the highest refinement levels, the gas structure of the halo—particularly in the medium-density regime—remains largely unaffected by the resolution ratio. Therefore, we adopt the highest resolution run of 0.1 \Ms\ for our present study. Despite these limitations, the resulting density profiles in the convergence tests show good agreement across different gas resolutions. Furthermore, we adopt gravitational softening lengths of $0.05$ pc for gas particles and $10,\mathrm{pc}$ for dark matter particles. Convergence tests in which the gas softening length is varied by a factor of ten in both directions show no significant differences in the final gas density profiles. 

\begin{figure}[tbh]
\centering
\includegraphics[width=0.6\columnwidth]{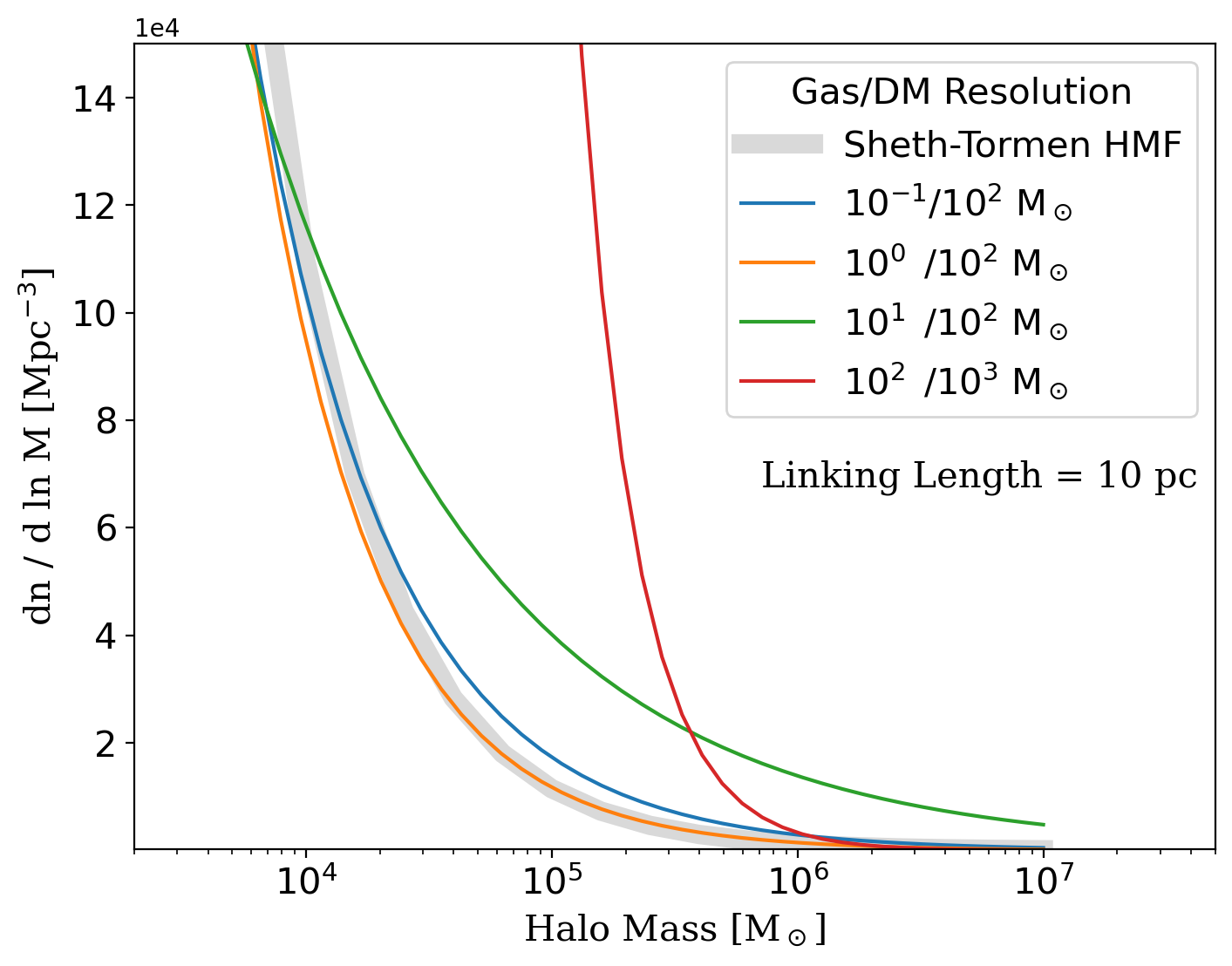}
\caption{Subhalo mass functions for a $10^7$ \Ms\ minihalo simulated using the particle-splitting method. The profiles show results from simulations with different gas and dark matter mass resolutions. Due to the limited number of halos in the mass range $10^3$–$10^7$ \Ms\ in our small-volume simulation, the HMF profiles are constructed from best-fit curves to the binned HMF data. The profiles converge at high resolution and agree well with the theoretical Sheth–Tormen model.} 
\label{fig:Halo_Mass_Func}
\end{figure}

Furthermore, we apply a friends-of-friends (FOF) algorithm \citep{More_2011} to the same dataset to identify substructure halos, using a linking length of 10 pc, and present the resulting halo mass function (HMF) in Figure~\ref{fig:Halo_Mass_Func}. The profiles of HMF start to converge for high-resolution runs and they match well with the standard $\Lambda$CDM halo mass function (e.g., the Sheth--Tormen HMF; \citealt{Sheth_1999}).

\subsection{Clump-Finding Algorithm}
Similar to present-day star formation, strong turbulence can typically produce clumpy structures of dense gas that host star formation. To quantify the clumpy structures in our simulations, we identify gas clumps within the central regions of the minihalos using the clump finder module in the \texttt{YT} analysis toolkit \citep{YT}, a Python-based platform for volumetric data visualization and analysis. The clump-finding algorithm, originally developed by \citet{bsmith09} and \citet{turk09}, first maps the gas particles onto an uniform grid and identifies structures by iteratively tracing isodensity contours, starting from a selected high-density region referred to as the master clump.
A clump is defined as a region enclosed by the lowest isodensity surface that contains a local density maximum, following the method described by \citet{Truelove1998}. However, only the gravitationally bound clumps are considered viable sites for Pop~III star formation. For the identified dense clumps, we reapply the clump-finding algorithm to identify gravitationally bound substructures
containing a minimum of five gas cells\footnote{Because the grid cell used in clump finding is finer than the particles, the resulting cell masses are smaller than the particle masses.} that satisfy the Jeans criterion.

\section{Primordial gas in mini halos}
Our simulations include a sample of 15 minihalos, each exhibiting distinct evolutionary behaviors. The detailed properties of each minihalo at the time of analysis are summarized in Table\ref{tab:results}.

\begin{table}[ht]
    \centering
    \begin{tabular}{c | c c c c c}
        \hline
        \textbf{Halo} & Virial Radius [pc] & Virial Mass [$10^5$ \Ms] & Gas [$10^5$ \Ms] & Dark Matter [$10^5$ \Ms] & Time [Myr] \\
        \hline
            \textbf{A} & 121.79 & 4.90 & 0.62 & 4.28 & 17.41 \\
            \textbf{B} & 133.74 & 6.37 & 0.87 & 5.49 & 17.71 \\
            \textbf{C} & 162.23 & 11.10 & 1.38 & 9.73 & 19.97 \\
            \textbf{D} & 161.42 & 11.54 & 1.16 & 10.38 & 18.61 \\
            \textbf{E} & 180.83 & 14.01 & 1.43 & 12.58 & 24.80 \\
            \textbf{F} & 196.59 & 18.90 & 1.75 & 17.15 & 22.13 \\
            \textbf{G} & 202.43 & 22.54 & 2.18 & 20.36 & 17.11 \\
            \textbf{H} & 213.30 & 25.43 & 2.74 & 22.68 & 18.91 \\
            \textbf{I} & 219.86 & 28.51 & 2.81 & 25.69 & 14.17 \\
            \textbf{J} & 253.11 & 30.81 & 2.98 & 27.83 & 14.75 \\
            \textbf{K} & 235.86 & 32.72 & 2.25 & 30.46 & 21.20 \\
            \textbf{L} & 245.31 & 37.20 & 4.80 & 32.40 & 19.21 \\
            \textbf{M} & 252.61 & 45.87 & 4.25 & 41.62 & 14.17 \\
            \textbf{N} & 282.98 & 56.85 & 4.13 & 52.72 & 20.28 \\
            \textbf{O} & 311.67 & 77.08 & 6.61 & 70.47 & 21.51 \\
        \hline
    \end{tabular}
    \caption{Physical properties of minihalos at the end of the simulations.} 
    \label{tab:results}
\end{table}

\subsection{Physical properties of primordial star-forming cloud}

During the formation of minihalos, the gravitational potential well formed by dark matter traps the surrounding primordial gas and concentrates it in the halo centers. Due to the tidal force from dark matter and inhomogeneous accretion, the accreting gas flow becomes highly anisotropic. Multiple accretion streams are injected into the center of the halo and drive a turbulent gas cloud. 

\begin{figure}
    \centering
    \includegraphics[width=\textwidth]{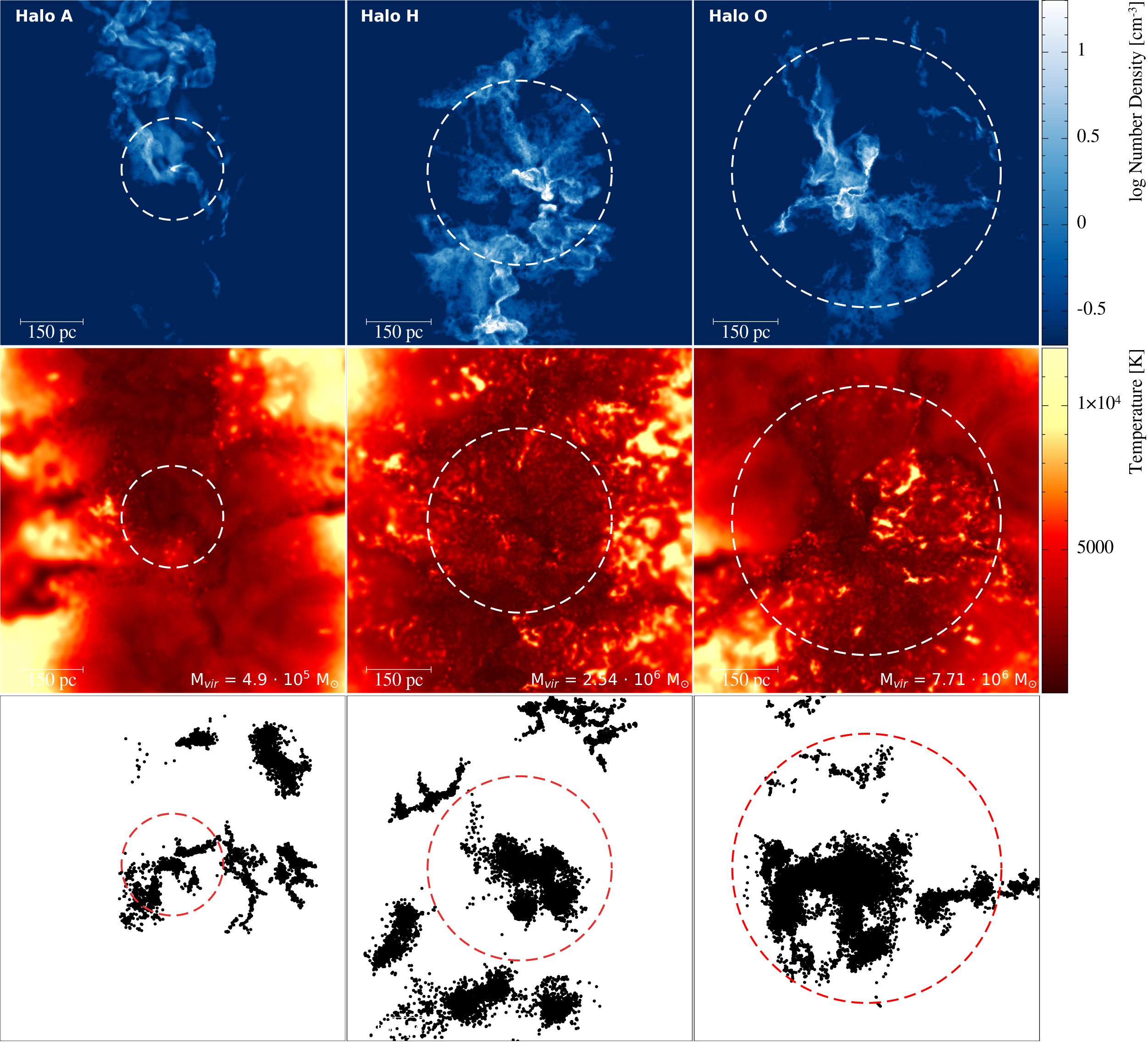}
    \caption{Gas density, temperature, and distribution of DM particles of \textbf{A, H,} and \textbf{O} halos at the end of the simulations. The bright color represents the high-dense region. The dashed circles show the virial sphere of the halo which increase with the halo mass. Fragmental structure appears at the halo centers, implying the turbulence formation due to gas accretion. The highest density region in our model, as shown in \ref{fig:Density_Distribution} could reach $10^{10}$ \NumDen. For comparison purpose, we set the density range to 0.3 - 33 \NumDen for highlighting the distribution of the high-density area.}
    \label{fig:Density_Temperature_and_DM_Distribution}
\end{figure}

We select three models including \textbf{Halo A} ($4.90 \times 10^5\,\Ms$), \textbf{Halo H} ($2.54 \times 10^6\,\Ms$), and \textbf{Halo O} ($7.71 \times 10^6\, \Ms$) to present small, medium and large halo masses for comparison. We first show the density slice of these models at the end of the simulations in Figure \ref{fig:Density_Temperature_and_DM_Distribution}. The gas distribution inside the halos looks highly complicated with the filamentary structure. 
Gravitationally driven supersonic turbulence triggers the formation of fragmented structures within the halo. The turbulence is driven on large scales, with a characteristic driving scale of approximately $\sim 3\,R_{\rm vir}$ based on \cite{2025_chen}. As a result, filamentary structures can form slightly beyond the virial radius; however, filaments in the outer halo remain significantly more diffuse than those in the central regions.

\begin{figure}
    \centering
    \includegraphics[width=\textwidth]{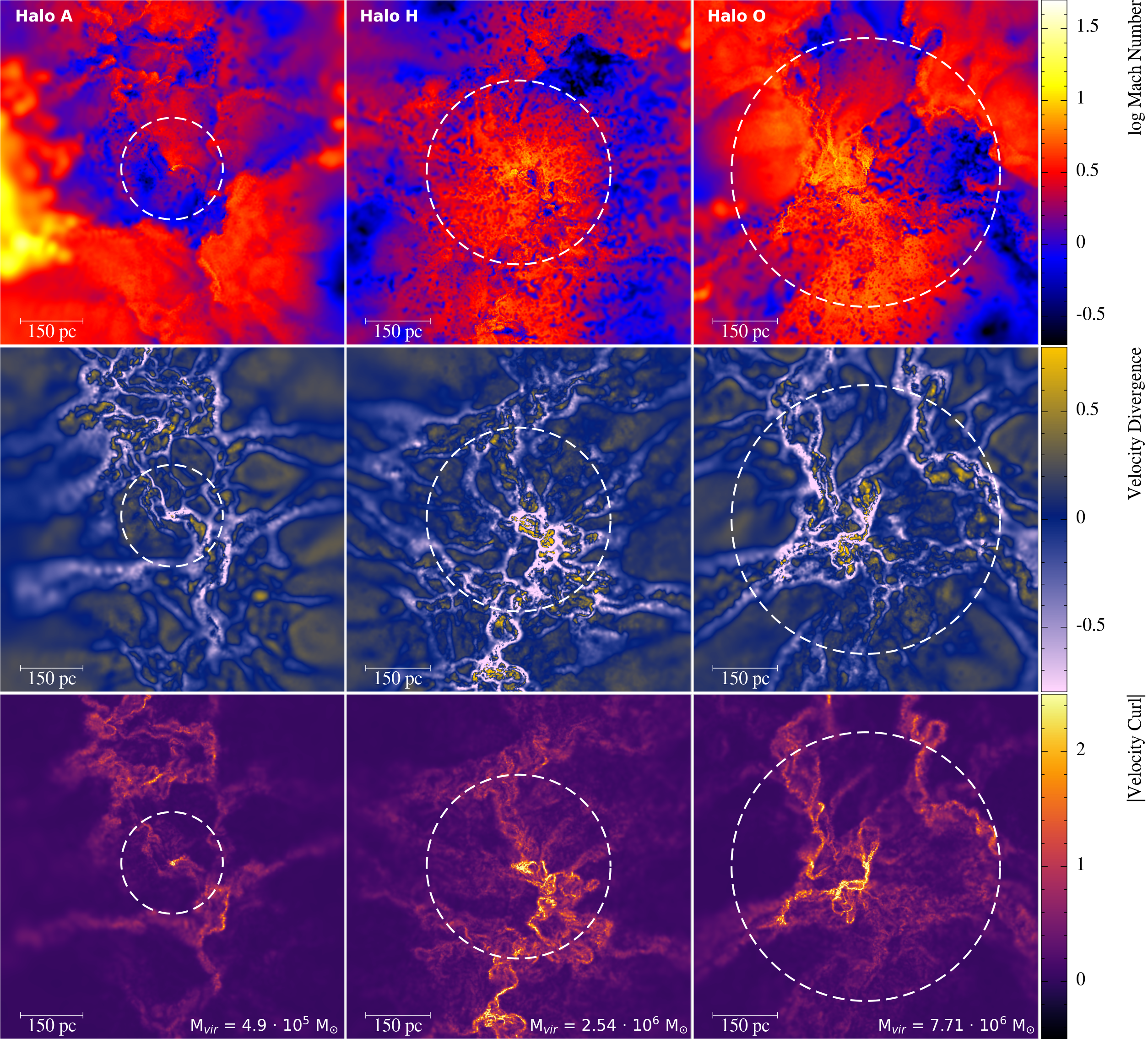}
    \caption{\Mach, velocity divergence and curl of \textbf{A, H,} and \textbf{O} halos at the end of the simulations. Overlapped area of the divergence and curl shows the highly turbulent motion of gas. The gas in the halo is dominated by supersonic motions with \Mach\ greater than unity, and the strength of the turbulence increases with halo mass. The curl and divergence of the velocity field further reveal the structure and dynamics of the accretion flows and are strongly correlated, particularly in the central regions of the halos.}
    \label{fig:Halo_Mach_Divergence_and_Curl}
\end{figure}

In Figure \ref{fig:Density_Temperature_and_DM_Distribution}, we also show the corresponding temperature distribution. The high-dense regions normally correspond to the low-temperature regions. The average temperature of the halos is $\sim 1000-2000$~K, and the center is a few hundred K, which is caused by molecular hydrogen cooling. The high-density region can form molecular hydrogen more efficiently and cool to a lower temperature. Furthermore, we present the flow dynamics—characterized by characteristic Mach numbers (\Mach), velocity divergence, and velocity curl for \textbf{Halos A}, \textbf{H}, and \textbf{O} in Figure~\ref{fig:Halo_Mach_Divergence_and_Curl}. Supersonic gas motions are predominantly confined within the halos. The presence of strong velocity divergence and curl demonstrates highly non-uniform gas accretion during minihalo assembly. The pronounced spatial correlation between regions of large divergence and curl indicates vigorous turbulent motions, particularly in the central regions of the halos.
 
\begin{figure}
   \centering
    \includegraphics[width=\textwidth]{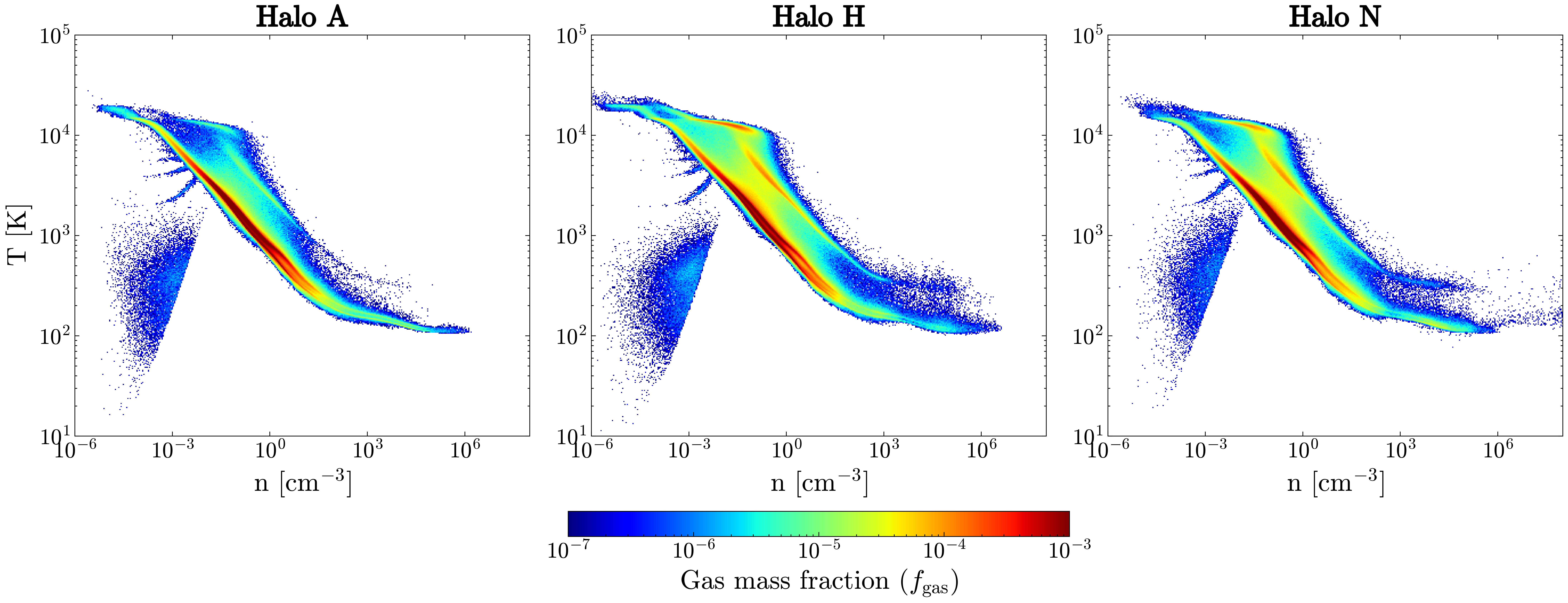}
\caption{Gas temperature–density phase diagram at the end of the simulations. The gas spans a density range from $10^{-5}$ to $10^{6}$~\NumDen, with corresponding temperatures from $\sim 10$ to $10^4$~K. In the low-density regime ($10^{-5} < n < 10^{-2}$~\NumDen), the gas temperature decreases with density due to adiabatic expansion. At higher densities ($n > 10^{-2}$~\NumDen), efficient radiative cooling by molecular hydrogen (H$_2$) and deuterated hydrogen (HD) causes the temperature to decrease as density increases. A distinct component of warm gas with $T \sim 10^4$~K is also present, originating from shock-heated, low-density gas in the outer regions of the halos.}
    \label{fig:Phase Diagram}
\end{figure}

The phase diagrams of \textbf{Halo A}, \textbf{H}, and \textbf{O} are shown in Figure~\ref{fig:Phase Diagram}. A prominent feature in all three models is the broad distribution of gas particles spanning from the low-density, high-temperature regime to the high-density, low-temperature regime. This trend reflects the primordial gas cooling processes included in our models. Initially, the simulations begin with diffuse and warm gas. As the halos evolve, gas accretes toward the central regions of the minihalos. With increasing density, molecular hydrogen (H$_2$) cooling becomes increasingly efficient, resulting in a gradual temperature decline at high densities. This cooling continues until the gas reaches the temperature floor set by H$_2$ and HD cooling, typically around $100$–$200$~K, depending on the local gas density \citep{Glover_2008}. However, the phase diagrams also include low-density gas outside the halo’s virial radius, where the temperature floor is set by the cosmic microwave background (CMB), approximately 50–60 K at $z \sim 18-20$. This dilute gas appears in the phase diagram at densities ($n \lesssim 10^{-3}$ \NumDen) and temperatures below 100 K.

The distant sharp boundary in the phase diagram corresponds to gas at low temperatures ($\sim 10^2$–$10^3$ K) and low densities ($\sim 10^{-4}$–$10^{-1}$ \NumDen). A similar feature was reported by \cite{Katz1996ApJS..105...19K}. This sharp boundary arises from the gas equation of state: during adiabatic compression, the temperature scales as $T \propto \rho^{\gamma-1}$, where $\gamma$ is the adiabatic index ($P \propto \rho^\gamma$); for an ideal monatomic gas, $\gamma = 5/3$.

\begin{figure}
    \centering
    \includegraphics[width=\textwidth]{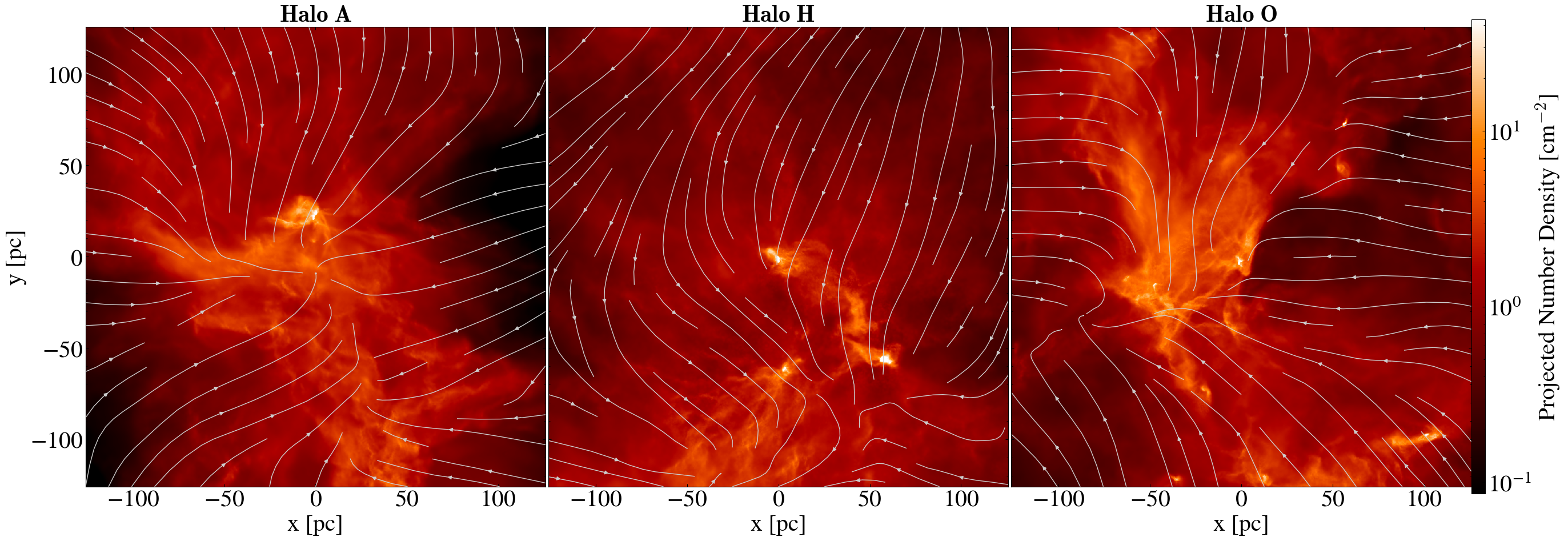}
    \caption{Gas accretion accretion onto minhalos. Stream lines represent the flow pattern of accreting gas. These lines converge to dense clumps at the halo center and stir up the turbulence gas.}
    \label{fig:Stream_Line}
\end{figure}

\subsection{Supersonic turbulence of primordial gas}

We examine the gas kinematics in the minihalos by presenting the gas density distribution and velocity streamlines in Figure~\ref{fig:Stream_Line}. The streamlines originate from large-scale gas accretion beyond the halo virial radius and exhibit complex, twisted patterns that converge toward the halo centers. This morphology indicates the presence of strong turbulent and convergent flows within the halos.

\subsection{Kinetic Energy power spectrum}

\begin{figure}
    \centering
    \includegraphics[width=0.6\textwidth]{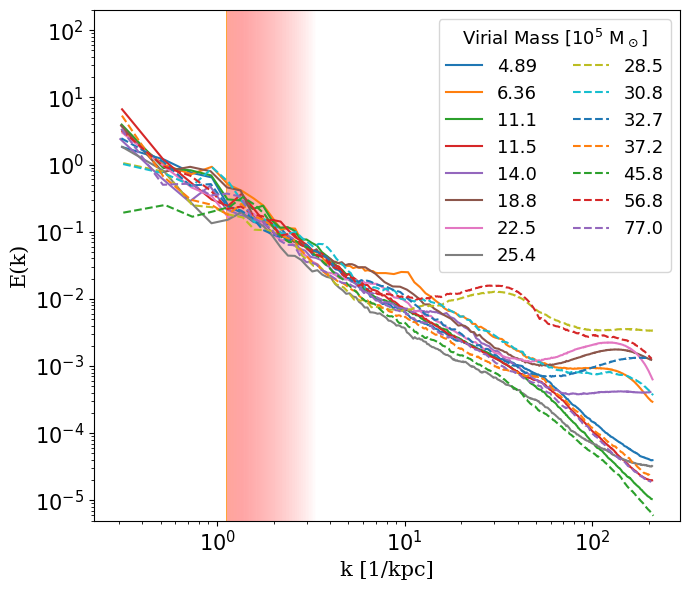}
    \caption{Kinetic energy spectra of halo gas for all halos. The x-axis shows the wave number in units of [kpc$^{-1}$], and the y-axis indicates the normalized kinetic energy of the gas. The red-shade region of $k \approx 1-3$ marginally represents the driving scale of turbulence. All curves follow closely to the Kolmogorov spectrum following $k^{-5/3}$, which closely matches the computed spectra from the halo gas, confirming the presence of turbulence in the halo environment.}
    \label{fig:Kolmogorov_Sepctrum}
\end{figure}

To quantify the flow kinematics within the halo, we show the gas kinetic energy power spectra of models \textbf{A}, \textbf{H}, and \textbf{O} in Figure~\ref{fig:Kolmogorov_Sepctrum}. The normalized spectra exhibit a power-law decay with a slope broadly consistent with the Kolmogorov spectrum \citep{1991_Kolmogorov, Shukurov_TurbulenceIntro}. The classical Kolmogorov spectrum can be divided into three characteristic regimes—from large to small scales: the energy injection (driving) scale, the inertial range, and the dissipation scale \citep{1991_Kolmogorov, Wang_2024}.

We perform this analysis for all models, with results summarized in Figure~\ref{fig:Kolmogorov_Sepctrum}, where the spectra are normalized to comparable amplitudes for easy comparison. Overall, most models closely follow the Kolmogorov slope, particularly on large scales. To identify the driving scale, $k_{\rm drive}$, we locate the local minimum in the power spectrum, defined as the point showing the largest relative decrease compared to the preceding point when moving from low to high wavenumber $k$. We find $k_{\rm drive} \sim 1$–$3~\mathrm{kpc^{-1}}$, corresponding to physical scales of $\sim 300$–$1{,}000$~pc. This scale separates the energy-injection regime from the inertial range. Given that the typical virial radius of our minihalos is $R_{200} \approx 100$–$300$~pc (Table~\ref{tab:results}), the largest turbulent eddies span $\sim 3$–$5\,R_{\rm vir}$, consistent with previous estimates that the turbulence driving scale is $\sim 3\,R_{\rm vir}$ \citep{2008_cuesta, 2023_gao, 2025_chen}. This corresponds to the so-called depletion radius, marking the outer boundary of the orbiting halo component. The region from the halo center out to $\sim 3\,R_{\rm vir}$ is therefore of particular interest, as it encompasses complex dynamics arising from interactions between infalling material and splashback components of both gas and dark matter, which naturally drive turbulence on halo-wide scales. Our highest spatial resolution reaches 0.05 pc, compared to $\sim 100$–200 pc in the original TNG50 simulations. Therefore, this driving scale is well resolved in our simulations but only marginally resolved in the original TNG50. In our zoom-in setup, splitting the parent gas and dark matter particles from the TNG50 simulation may introduce numerical artifacts and affect the effective driving scale inferred in our simulations, particularly in smaller halos. However, for larger mini halos with virial radius $\geq 200$ pc, the results should remain robust, as their corresponding driving scales ($\geq 600$ pc) are resolved by approximately tens of gas and dark matter particles in the TNG50. 

On smaller scales ($\lesssim 30$~pc), some models exhibit excess power relative to the Kolmogorov expectation, likely due to small-scale dark-matter substructures that enhance gas accretion and increase local gas velocities. The eventual dissipation of turbulence occurs at much smaller scales \citep{yuen2022turbulentuniversalgalactickolmogorov}, which are not resolved in our simulations. Instead, turbulence is dissipated through numerical viscosity at the particle-mesh level.

\subsection{Supersonic Turbulence}

\begin{figure}
    \centering
    \includegraphics[width=0.6\linewidth]{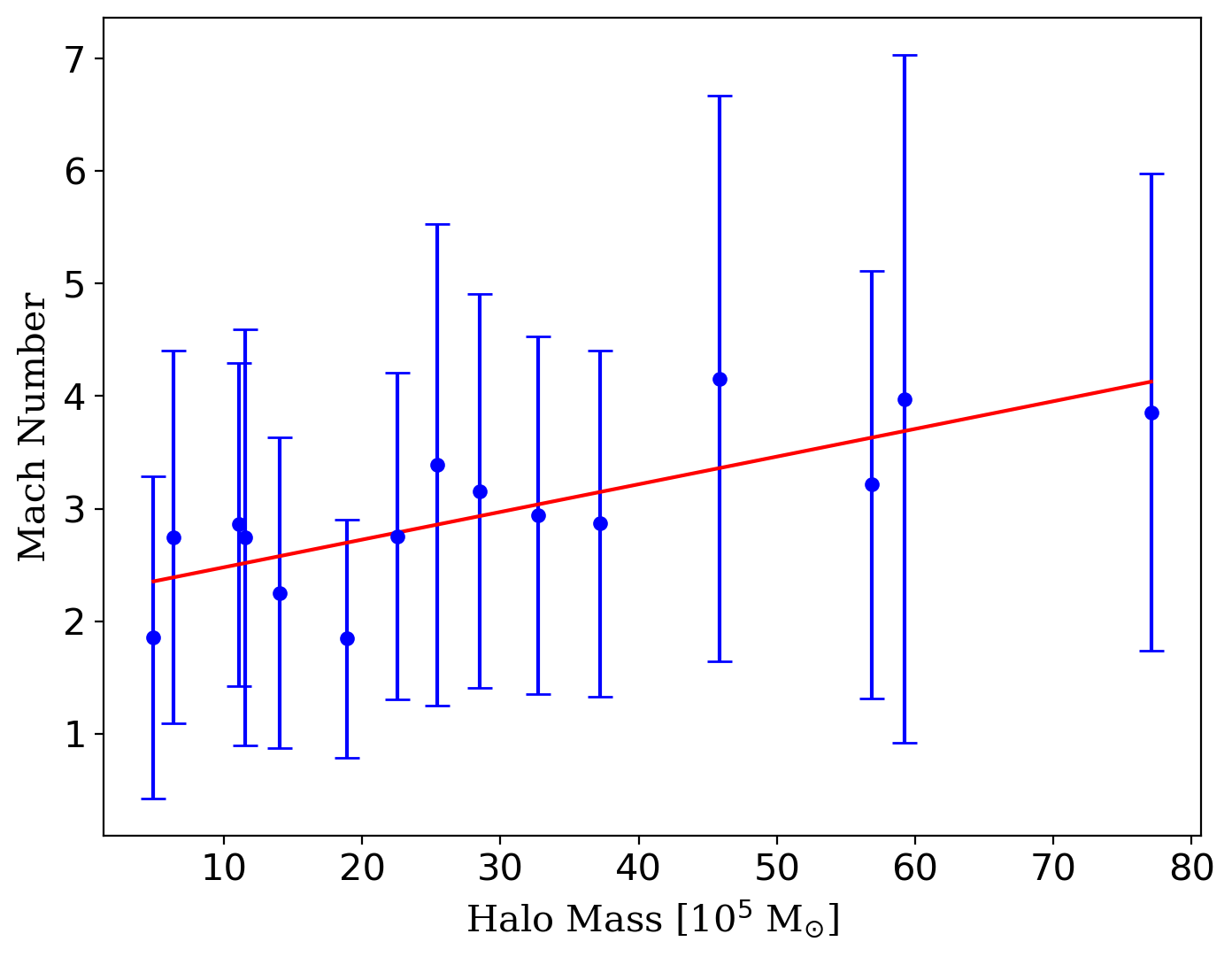}
    \caption{ The relation between the mean \Mach\ and its halo mass. The mean \Mach\ ranges from 1.8 to 3.8, roughly increasing with the halo mass. The read line shows the best fit profile for the data points, which the average \Mach\ increases by one as the halo mass increases by $2.8\times 10^6\ \Ms$. The error bars are derived from the distribution of \Mach\ of the gas particles. The vertical lengths of the error bars represent one standard deviation from the mean value.}
    \label{fig:HaloMass_to_Mach_fit}
\end{figure}

We now examine the characteristic turbulent \Mach\ in our simulations. The \Mach\ number quantifies the ratio of the gas velocity to the local sound speed and is defined as \Mach~$\equiv v / c_s$, where $v$ is the gas velocity relative to the bulk motion of the halo. The local sound speed, $c_s$, is estimated as

 \begin{equation}
 c_s = \sqrt{\frac{\gamma R T}{\mu m_{\rm H}}} ,
 \end{equation}

where $\gamma$ is the adiabatic index, $R$ is the ideal gas constant, $m_{\rm H}$ is the hydrogen mass, and $\mu$ is the mean molecular weight of the gas. In general, the sound speed depends on both the gas temperature and the electron abundance. We focus our analysis on the gas within each halo and present the relationship between \Mach\ and halo mass in Figure~\ref{fig:HaloMass_to_Mach_fit}. Our results reveal a positive correlation between the characteristic \Mach\ and the halo's virial mass. Specifically, \Mach\ increases with increasing halo mass, with typical values ranging from \Mach $\approx 1.9$ to $4.1$ for halos with masses between $4.9 \times 10^5\,\Ms$ and $7.7 \times 10^6\,\Ms$. The error bars indicate a scatter of \Mach\ = $1-3$, depending on the specific halo. This trend suggests that more massive minihalos generate stronger turbulent flows, likely due to deeper gravitational potentials and higher infall velocities.

\begin{figure}
    \centering
    \includegraphics[width=0.6\linewidth]{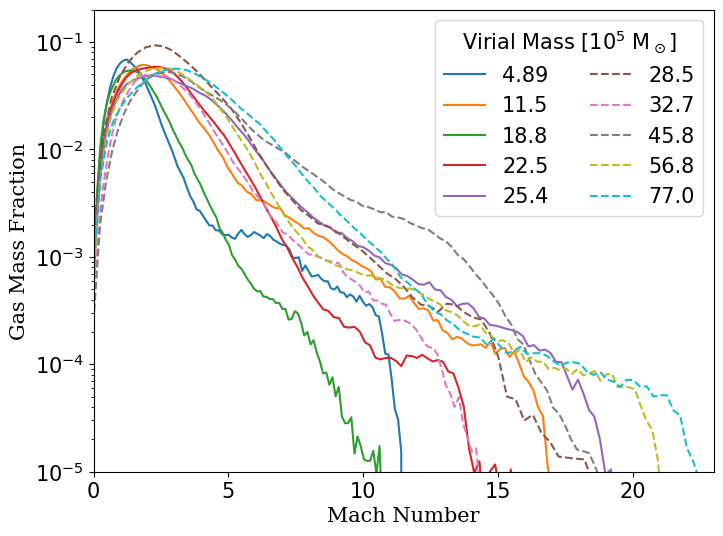}
    \caption{Gas mass fraction as a function of \Mach. The distribution follows an approximately log-normal shape for \Mach$ < 10$, with an extended high-\Mach\ tail reaching values of \Mach$ \sim 10-35$. As the halo mass increases, both the peak of the distribution and the extent of the high-\Mach\ tail shift toward higher \Mach, indicating stronger turbulence in more massive minihalos.
 }
    \label{fig:Mach_Distribution}
\end{figure}

Furthermore, we show the gas mass fraction as a function of the \Mach\ in Figure \ref{fig:Mach_Distribution}.
While the overall distribution appears similar across different halos, we note that the maximum \Mach\ in model \textbf{N} reaches up to $\sim 24$, whereas most other models exhibit peak \Mach\ values below 24. The characteristic \Mach\ number — corresponding to the peak of the curves in Figure \ref{fig:Mach_Distribution} — lies within the range of 3–5 for all models. Interestingly, the high-\Mach\ tail extends further for more massive halos. About $0.01$–$0.1\%$ of the gas can reach \Mach\ $\geq 10$. Given a total halo gas mass of $\sim 10^6\ \Ms$, this implies that roughly $1000\ \Ms$ of gas can reach such high \Mach\ numbers.

\begin{figure}
    \centering
    \includegraphics[width=0.6\linewidth]{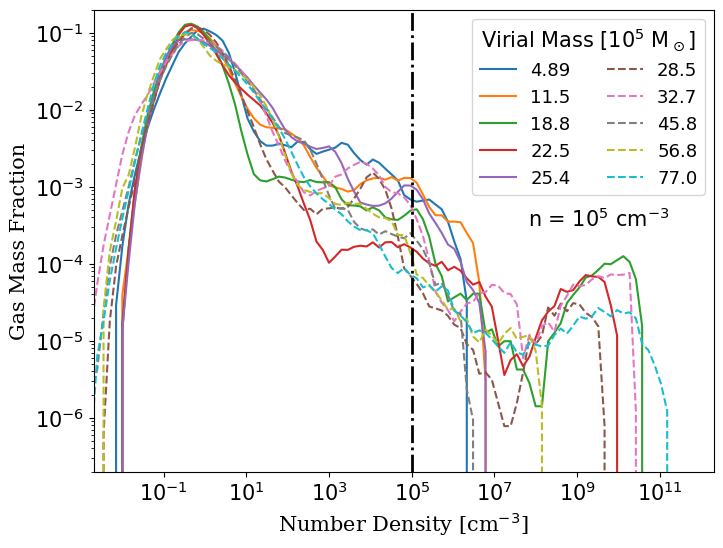}
    \caption{Gas mass fraction as a function of its number density. A vertical dot-dashed line marks $n = 10^{5}$~\NumDen, commonly considered the critical density for the gravitational collapse of the first star-forming gas. For densities below $n \lesssim 10^{2}$~\NumDen, the mass fraction profiles are remarkably similar across all halos. However, at higher densities ($n > 10^{2}$~\NumDen), the profiles begin to diverge significantly, reflecting varying degrees of dense gas formation and collapse in different minihalos.
}
    \label{fig:Density_Distribution}
\end{figure}

Figure \ref{fig:Density_Distribution} shows the gas mass fraction as a function of number density. We find a consistent trend across all models: the peak mass fraction is located at $n\approx 1$ \NumDen, corresponding to the inner halo region. A vertical line marks $n = 10^5$ \NumDen, the typical density threshold for the primordial gas to collapse into Pop~III stars \citep{op01,Bromm_2002,chen15}. All models exhibit varying fractions of gas exceeding this threshold, indicating that star-forming conditions are achieved to different extents across the halos.

\subsection{Clumpy structures of the first-star clouds}
\begin{figure}
    \centering
    \includegraphics[width=\textwidth]{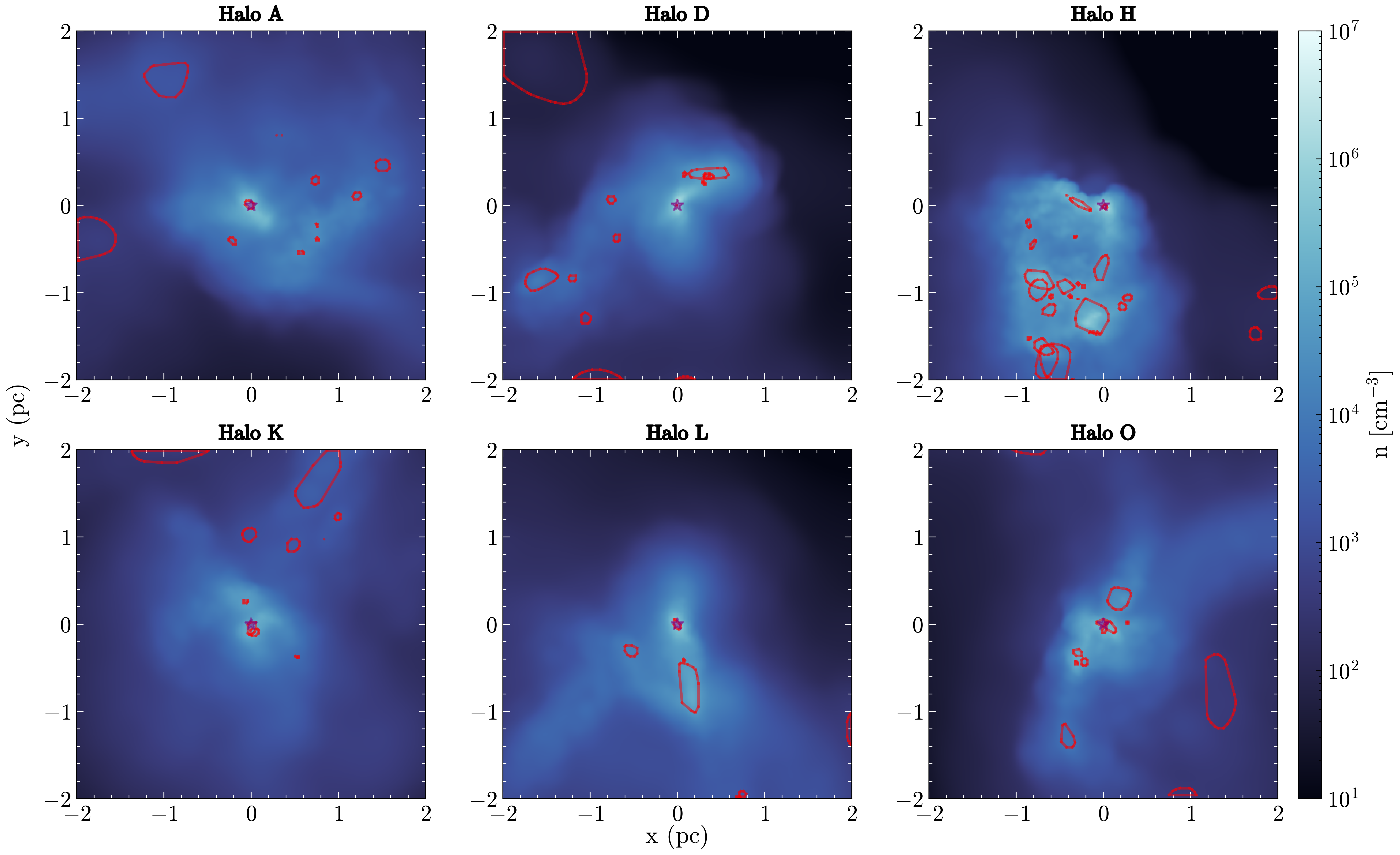}
\caption{Clumpy structures within the central regions of the minihalos. Red contours trace gas clumps that emerge from turbulent flows. While many of these clumps are not yet gravitationally bound, they may become bound over time through continued gas accretion. At the current stage of the simulation, only the dense clump marked with red stars—located at the halo center—has reached the threshold for gravitational binding, and its has begun to collapse due to Jeans instability.
}
\label{fig:clump_0.1virial_projection_contour}
\end{figure}

We use a clump-finding approach to identify the number, morphology, and mass of star-forming clouds. Figure \ref{fig:clump_0.1virial_projection_contour} shows the final snapshot of the clump distributions based on this method. Multiple gas clumps are found in each halo, and at this stage, one gravitationally bound clump (dense clump) has formed for every halo. These dense clumps are undergoing Jeans instability and are expected to collapse imminently to form stars.

The mass of dense clumps provides an upper limit for the mass of the Pop~III star that may form within it. Therefore, both the mass and the number of dense clumps set important constraints on the characteristic mass scale and the initial mass function (IMF) of Pop~III stars. We summarize the physical properties of these dense clumps in Table~\ref{tab:clump_properties}. Across all halos, dense clump masses range from 2.6~\Ms\ to 66.5~\Ms, with central densities from $4.8 \times 10^5$ to $3.81 \times 10^9$ \NumDen, and temperatures between 111–321 K. All dense clumps exceed their corresponding Jeans masses, confirming their gravitational instability.

Once a dense clump begins to collapse, the simulation timestep rapidly decreases—by a factor of $\sim10^4$—which effectively halts the simulation. To follow the collapse further, higher resolution or the introduction of sink particles is necessary. Some clumps visible in Figure \ref{fig:clump_0.1virial_projection_contour} may eventually evolve into dense clumps if the simulations are allowed to proceed longer.
\begin{table}[ht]
{
    \centering
    \begin{tabular}{c | c c c c c}
        \hline
        Halo & $r_\mathrm{eff}$ [pc] & $M_\mathrm{clump}$ [\Ms] & $M_\mathrm{J}$ [\Ms] & $n$ [cm$^{-3}$] & $T_\mathrm{C}$ [K] \\
        \hline
        A  & 0.049  &  13.79   &  2.99  &  $1.14 \times 10^6$  & 111.44 \\
        B  & 0.024  &  12.58   &  0.29  &  $2.30 \times 10^9$  & 294.57 \\
        C  & 0.050  &  11.96   &  3.50  &  $7.62 \times 10^5$  & 108.35 \\
        D  & 0.066  &  55.97   &  3.02  &  $2.23 \times 10^6$  & 140.46 \\
        E  & 0.035  &  50.21   &  0.62  &  $7.56 \times 10^7$  & 158.22 \\
        F  & 0.024  &  28.85   &  0.23  &  $3.44 \times 10^9$  & 294.68 \\
        G  & 0.031  &  66.50   &  0.52  &  $8.98 \times 10^8$  & 321.84 \\
        H  & 0.019  &   2.62   &  2.37  &  $2.90 \times 10^6$  & 130.31 \\
        I  & 0.024  &   8.12   &  0.41  &  $4.81 \times 10^8$  & 222.83 \\
        J  & 0.024  &   8.48   &  0.24  &  $3.81 \times 10^9$  & 308.16 \\
        K  & 0.059  &   7.89   &  5.72  &  $3.13 \times 10^5$  & 111.67 \\
        L  & 0.047  &  12.35   &  3.16  &  $1.11 \times 10^6$  & 114.71 \\
        M  & 0.019  &  10.36   &  0.87  &  $2.03 \times 10^7$  & 128.33 \\
        N  & 0.028  &  61.96   &  0.81  &  $1.47 \times 10^8$  & 235.78 \\
        O  & 0.063  &  13.31   &  5.06  &  $4.48 \times 10^5$  & 116.04 \\
        \hline
    \end{tabular}
    \caption{Physical properties of the dense clump in each halo.   The table lists clump size ($r_\mathrm{eff}$), mass ($M_\mathrm{clump}$), jeans mass ($M_\mathrm{J}$), and peak density ($\rho_\mathrm{max}$).  $r_\mathrm{eff}$ is defined by $(3V_\mathrm{c}/4\pi)^{1/3}$, where $V_\mathrm{c}$ is the volume of the clump. $M_\mathrm{J}$ here is estimated according to Equation 2 in \cite{2024Chen}. These clumps exceeding their jeans should collapse and form into stars shortly. Note: $M_\mathrm{clump}$ and $M_\mathrm{J}$ are calculated from the gas cell masses defined by the clump-finding algorithm, which uses a grid finer than the original particle sampling. Because of this higher resolution, the gas cell masses are smaller than the gas particle masses, allowing both quantities to achieve a higher effective mass resolution than the original gas particles.}
    }

    \label{tab:clump_properties}
\end{table}

\begin{figure}
    \centering
    \includegraphics[width=\textwidth]{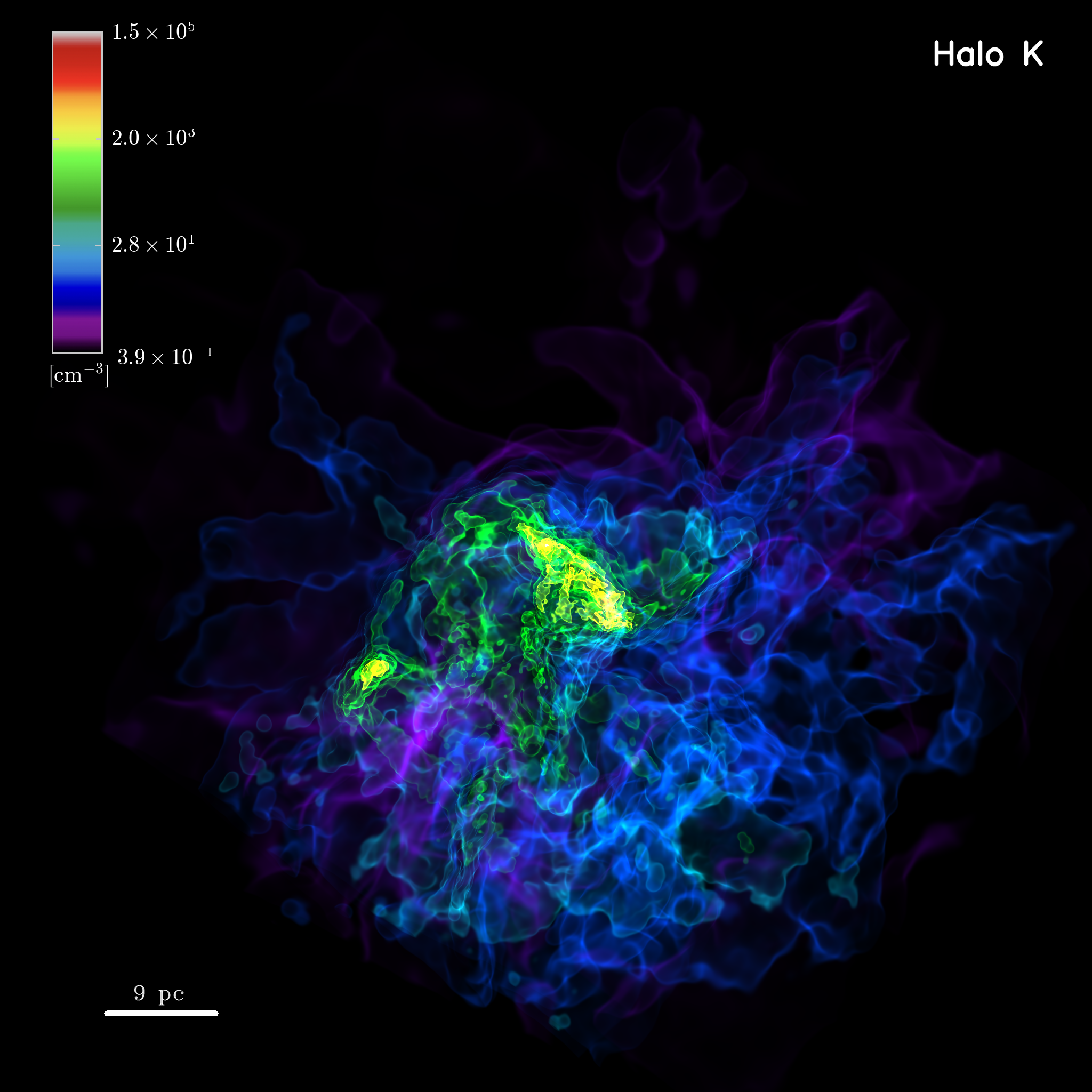}
    \caption{3D gas density structure within $0.1\,R_{\rm vir}$ of \textbf{Halo K}, highlighting two distinct clumps (visible as separate yellow-green nuggets). The detailed substructures within these clumps illustrate the influence of supersonic turbulence in the first star-forming region. Both clumps are expected to grow further through continued gas accretion and eventually collapse to form stars. The final mass of each clump sets an upper limit on the mass of the stars that may form within them.
 }
    \label{fig:clump_volume}
\end{figure}

\section{Discussion}

Previous studies were unable to resolve and follow the formation of primordial turbulence during the early assembly of minihalos. \citet{abel1998formationfragmentationprimordialmolecular_PopIII_0, Bromm_2013_The_first_star} suggested that turbulence within the first star-forming clumps is subsonic or transonic and plays a limited role in star formation. However, these earlier works typically employed hierarchy zoom-in techniques using AMR or SPH, with the highest-resolution only for a small regions at center of the minihalos. As a result, they lacked sufficient resolution on larger scales to capture the full development of the accretion flows and turbulence found in \cite{2025_chen}.

Our simulations successfully resolve the formation and evolution of turbulence during minihalo assembly. This turbulence is driven primarily by gravitational potential energy released during gas accretion onto the minihalos. We find that most of the halo gas become supersonic with a characteristic \Mach\ of ~$\sim 2$--$5$, consistent with those adopted in driven turbulence simulations by \citet{tang2024clumpy}. In some halos, a small fraction of gas ($\sim$0.01--0.1\%) can even reach \Mach~$> 10$. We also find a positive correlation between the characteristic \Mach\ and the minihalo mass.

Each minihalo contains approximately $10^2$–$10^3\ \Ms$ of dense gas ($n \geq 10^5$ \NumDen) that is expected to collapse and form Pop~III stars. This range of available gas mass is broadly consistent with previous studies \citep{abel1998formationfragmentationprimordialmolecular_PopIII_0, Bromm_2013_The_first_star}. However, the supersonic turbulence present in our simulations breaks spherical symmetry, leading to the formation of clumpy gas structures (Figure~\ref{fig:clump_0.1virial_projection_contour}). In several halos (e.g., \textbf{H} and \textbf{D}), multiple clumps are identified, although only one reaches the dense, gravitationally bound stage at the epoch shown.

The masses of these dense clumps span $2.6$–$66.5\ \Ms$ across the halo sample, setting an upper limit on the masses of individual Pop~III stars in the absence of subsequent clump mergers. The emergence of such clumps reduces the characteristic mass scale of Pop~III stars and represents a qualitative departure from the classical disk fragmentation scenario observed at AU scales in earlier studies \citep{turk09, clark11, Greif_2015, gen22_fragement}.

Figure~\ref{fig:clump_volume} illustrates the three-dimensional gas density structure within $0.1\ R_{\rm vir}$ of Halo \textbf{K}, where two distinct, elongated dense clumps are clearly visible at the halo center.

Our simulations are evolved with full physics for 14–24~Myr, which is shorter than the lifetimes and complete accretion histories of low-mass protostars. However, the primary goal of this study is to characterize the formation of gravitationally bound clumps and to quantify how supersonic turbulence regulates their mass spectrum, rather than to follow the subsequent evolution of individual protostars. In each simulation, we evolve the system until multiple clumps form and at least one dense clump exceeds the local Jeans mass. At this stage, clump masses are largely set by the turbulent velocity field, shock compression, and fragmentation driven by supersonic motions.

These clump masses therefore provide a physically meaningful upper limit on the stellar masses that can form, since a star cannot exceed the mass reservoir of its natal clump. We find a broad mass distribution of dense clumps, reflecting the stochastic nature of turbulence-driven fragmentation and offering insight into the characteristic mass scale of Pop~III star formation. Although lower-mass clumps may continue to accrete beyond the simulated time, their long-term growth is constrained by the finite gas reservoir, competition among neighboring clumps, and the global turbulent environment. Consequently, the clump masses identified in our simulations provide a robust estimate of the maximum mass available to individual Pop~III stars.

\subsection{Supersonic Turbulence from Streaming Velocity}

\citet{Tseliakhovich_2010} showed that relative streaming velocities between baryons and dark matter arise from baryon acoustic oscillations (BAOs) in the early Universe. These velocities can be supersonic and significantly affect Pop~III star formation by modifying the gas content and dynamical evolution of primordial minihalos \citep{Schauer_2021, Lake_2022, Hirano_2023, Hirano_2024, Hirano_2025}. Supersonic streaming velocities have been incorporated into simulations using a variety of theoretical treatments \citep{greif11, maio11, stacy11a, naoz12, naoz13, O_Leary_2012, latif14, Schauer_2019, Nebrin_2023, Hegde_2023}, leading to differences in the predicted strength and impact of the streaming motions. Depending on the treatment, streaming velocities can either enhance gas fragmentation and promote the formation of multiple Pop~III stars, or suppress fragmentation, allowing the collapse of massive gas clouds that may form supermassive stars.

In more massive primordial halos, such as atomic-cooling halos with masses $\gtrsim 10^9\ \Ms$, streaming velocities can facilitate the formation of supermassive stars that subsequently seed supermassive black holes \citep{Hirano_2017}. Recent work by \citet{Lake_2023, Lake_2024} further suggests that supersonically induced gas objects (SIGOs), formed due to baryon--dark matter streaming at recombination, may be progenitors of early globular clusters, with star formation rates increasing with streaming velocity.

In this study, we focus on minihalos with masses of $\sim 10^5$--$10^7\ \Ms$, assuming a region where the local streaming velocity is small and can be approximated as zero. This allows us to isolate the role of pure gas accretion during minihalo assembly. Nevertheless, if large-scale streaming velocities cascade down to minihalo scales, they could further amplify the supersonic turbulence identified in our simulations.

\subsection{Possible Observables}
Direct detection of individual Pop III stars remains beyond the capabilities of current facilities, including JWST, likely due to their short lifetimes, high redshifts, and intrinsic faintness. Nevertheless, the radiative and chemical feedback from the first stars and their supernovae is strongly mass dependent and leaves observable imprints on the earliest galaxies—key targets for JWST—by shaping their thermal, ionization, and chemical evolution. In addition, the abundance patterns of extremely metal-poor stars in the Milky Way and nearby dwarf galaxies retain the nucleosynthetic signatures of the first supernovae, providing an indirect probe of Pop III progenitor masses. Interpreting these observations requires a robust understanding of the initial mass function (IMF) of the first stars. Our study, which focuses on the formation and mass distribution of protostellar clumps regulated by supersonic turbulence, therefore provides an essential step toward constraining the Pop III IMF and linking theoretical models to observable signatures.

\subsection{Important Caveats}
External stellar feedback is not included in our simulations, consistent with our assumption that the selected minihalos reside in pristine environments. In general, additional feedback processes can either enhance or suppress filamentary structures in primordial clouds and may alter the Pop~III initial mass function. However, previous studies have shown that during the early stages of minihalo assembly, the impact of stellar feedback is relatively weak compared to the turbulence generated by gravitational accretion and halo assembly \citep{Wise_2007, turk09, Greif_2008, greif11, Latif_2013}.

Although minihalos may, in principle, be affected by radiative feedback—such as Lyman--Werner (LW) radiation—from nearby star-forming regions, we mitigate this effect by selecting halo samples with no pre-existing stars within the simulation volume. As a result, the LW background is expected to play only a minor role in our study. If LW radiation were included, its heating effect would increase the gas sound speed and reduce the \Mach, thereby weakening the level of supersonic turbulence \citep{Haiman_2000, Machacek_2001, Norman_2008}.

We also neglect baryon--dark matter streaming velocities associated with baryon acoustic oscillations in the present work. Such streaming motions would likely enhance turbulent motions during minihalo assembly and further strengthen the resulting supersonic turbulence \citep{Tseliakhovich_2010, greif11, Schauer_2017}. Finally, if our simulations were extended into the clump-collapse and star-formation phases, additional feedback mechanisms—such as protostellar radiation, jets, or stellar winds—would inject momentum into the surrounding gas and could further amplify supersonic turbulence, similar to what is observed in present-day star-forming environments \citep{Stacy_2016, Hirano_2014_PopIII_3, Hosokawa_2016_PopIII_4}.

\section{Conclusion}

We investigate the universality and diversity of supersonic turbulence in primordial halos by simulating the formation of 15 minihalos. Our \GIZMO\ simulations use initial conditions drawn from large-scale cosmological \TNG\ simulations, enhanced by a robust particle-splitting technique to assess turbulent structures within primordial gas clouds—the birthplaces of Pop~III stars. The results show that the turbulence arises naturally from gas accretion into the gravitational potential wells of dark matter during minihalo formation. Gas velocities in the minihalos span from subsonic to supersonic, reaching \Mach\ up to $\sim30$, clearly indicating the presence of supersonic turbulence. 

In the central regions of halos, turbulent gas exhibits significant compressibility and vorticity, coinciding with high-density zones where molecular hydrogen cooling becomes efficient. The characteristic \Mach\ of halos is between 2 and 5 and correlates positively with the mass of the halo. Supersonic turbulence disrupts spherical collapse, producing clumpy filamentary structures, consistent with predictions by \citet{tang2024clumpy}. The physical properties of primordial turbulence revealed by our simulations offer critical insights into the long-standing problem of Pop~III star formation.

Our findings show that the clump masses produced by supersonic turbulence significantly influence the characteristic mass scale of the first stars and their associated stellar feedback, which critically shape the physical properties of the first galaxies. These implications can be tested with via the first galaxy observations with the James Webb Space Telescope (JWST).

\acknowledgments
Authors thank Chi-Hung Lin, Po-Feng Wu, and Chorng-Yuan Hwang for their useful discussions. We also thank Ching-Yao Tang for his help with the data analysis. KC acknowledges the support of the Alexander von Humboldt Foundation and Heidelberg Institute for Theoretical Studies. This research is supported by the National Science and Technology Council, Taiwan, under grant No. NSTC 113-2112-M-001-028-,114-2112-M-001-012-, 114-2811-M-001-094, and the Academia Sinica, Taiwan, under a career development award under grant No. AS-CDA-111-M04. This research was supported in part by grant NSF PHY-2309135 to the Kavli Institute for Theoretical Physics (KITP) and grant NSF PHY-2210452 to the Aspen Center for Physics. Our computing resources were supported by the National Energy Research Scientific Computing Center (NERSC), a U.S. Department of Energy Office of Science User Facility operated under Contract No. DE-AC02-05CH11231 and the TIARA Cluster at the Academia Sinica Institute of Astronomy and Astrophysics (ASIAA).

\bibliographystyle{yahapj}
\bibliography{refs,refs_pop3,refs_extra}

\end{document}